\documentclass[sigconf, 10pt]{acmart} 

\usepackage{booktabs} 
\usepackage{epsfig}
\usepackage{endnotes}
\usepackage{xcolor}
\usepackage{subcaption}
\usepackage{makecell}
\usepackage{amsmath} 
\usepackage{pifont}

\usepackage{tikz}
\usetikzlibrary{arrows.meta, positioning, shapes, fit}

\def\BibTeX{{\rm B\kern-.05em{\sc i\kern-.025em b}\kern-.08em
    T\kern-.1667em\lower.7ex\hbox{E}\kern-.125emX}}

\acmYear{2026}\copyrightyear{2026}
\setcopyright{cc}
\setcctype[4.0]{by}
\acmConference[NOSSDAV '26]{Workshop on Network and Operating System Support for Digital Audio and Video}{April 4--8, 2026}{Hong Kong, Hong Kong}
\acmBooktitle{Workshop on Network and Operating System Support for Digital Audio and Video (NOSSDAV '26), April 4--8, 2026, Hong Kong, Hong Kong}
\acmDOI{10.1145/3798065.3798083}
\acmISBN{979-8-4007-2534-0/26/04}

\begin{document}

\title{HybridPrompt: Bridging Generative Priors and Traditional Codecs for Mobile Streaming}

\author{Liming Liu}
\author{Jiangkai Wu}
\affiliation{%
  \institution{Peking University}
  \country{} 
}

\author{Haoyang Wang}
\author{Peiheng Wang}
\affiliation{%
  \institution{Peking University}
  \country{} 
}

\author{Zongming Guo}
\author{Xinggong Zhang}
\affiliation{%
  \institution{Peking University}
  \country{} 
}



\begin{abstract}




In Video on Demand (VoD) scenarios, traditional codecs are the industry standard due to their high decoding efficiency. However, they suffer from severe quality degradation under low bandwidth conditions. While emerging generative neural codecs offer significantly higher perceptual quality, their reliance on heavy frame-by-frame generation makes real-time playback on mobile devices impractical. We ask: is it possible to combine the blazing-fast speed of traditional standards with the superior visual fidelity of neural approaches? We present \emph{HybridPrompt}, the first generative video coding system capable of achieving real-time 1080p decoding at over 150 FPS on a commercial smartphone. Specifically, we employ a hybrid architecture that encodes Keyframes using a generative model while relying on traditional codecs for the remaining frames. A major challenge is that the two paradigms have conflicting objectives: the "hallucinated" details from generative models often misalign with the rigid prediction mechanisms of traditional codecs, causing bitrate inefficiency. To address this, we demonstrate that the traditional decoding process is differentiable, enabling an end-to-end optimization loop. This allows us to use subsequent frames as additional supervision, forcing the generative model to synthesize keyframes that are not only perceptually high-fidelity but also mathematically optimal references for the traditional codec. By integrating a two-stage generation strategy, our system outperforms pure neural baselines by orders of magnitude in speed while achieving an average LPIPS gain of 8\% over traditional codecs at 200kbps.
\end{abstract}

\maketitle




\section{Introduction}

\begin{figure}[t]
    \centering
    \includegraphics[width=0.45\textwidth]{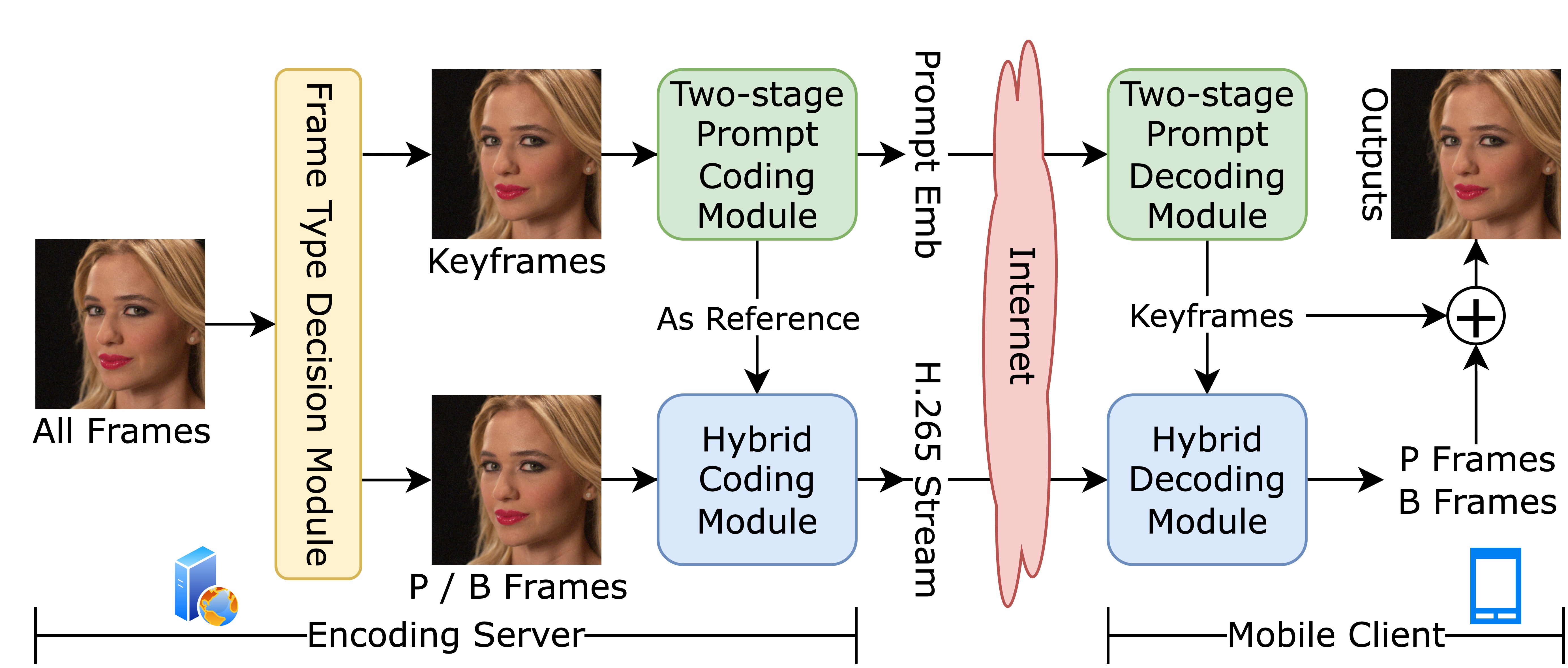}
    \vspace{-3mm}
    \caption{Overview of HybridPrompt.}
    \label{fig:overview}
    \vspace{-6mm}
\end{figure}

Video on Demand (VoD) services dominate global network traffic~\cite{global-phenomena24}, yet delivering high-quality streams to mobile users in unstable network environments remains a fundamental challenge. In scenarios with fluctuating bandwidth, such as crowded subways or rural areas, traditional video coding standards like H.264~\cite{264}, H.265~\cite{265}, and VP9~\cite{mukherjee2015technical} often fail to maintain acceptable visual quality. These standards work by dividing video into independent reference pictures called \textbf{Keyframes (I-frames)} and dependent frames called \textbf{Predicted frames (P-/B-frames)} that only record changes. Under strict bitrate constraints, the codec is forced to heavily compress the I-frames to save data. Since P- and B-frames rely entirely on I-frames for reconstruction, a low-quality, blurry I-frame causes error to propagate through the entire video segment, resulting in a poor viewing experience~\cite{ebenezer2022subjective}.

Generative Semantic Models offer a potential solution by reconstructing high-fidelity images from compact descriptions, but they currently lack the decoding speed required for mobile playback. The core mechanism of approaches like Promptus~\cite{wu2024promptus} relies on a "gradient based inversion": an offline encoding process that iteratively optimizes a tiny text-like code (prompt) that drives a model (e.g., Stable Diffusion~\cite{sd}) to synthesize an image. When encoding a frame, it adjust the small prompt-like code until the generated output matches the source video. This allows for incredibly low bandwidth usage since only the compact codes are transmitted. However, the tradeoff is the computational cost of decoding: existing generative video solutions typically require running this heavy generation process for \textit{every single frame}. On a mobile device with limited power, this approach is prohibitively slow, often taking seconds to generate just one frame, whereas smooth video requires at least 30 FPS.

Is it possible to combine the superior perceptual quality of generative models with the blazing-fast decoding speed of traditional codecs? We propose \emph{HybridPrompt}, a novel hybrid video coding strategy designed to achieve the best of both worlds. Our key insight is to use the heavy generative model \textit{only} for I-frames, while using the lightweight traditional codec for the motion-based P- and B-frames. This creates a "Bitrate Arbitrage": because the generative model can compress an I-frame to a negligible size while maintaining high visual quality, we save a massive amount of data. We then reallocate this saved bitrate budget to the traditional encoded P- and B-frames. As a result, the P- and B-frames receive enough data to maintain high clarity, and because they are decoded by the traditional hardware codec, the overall playback remains smooth and real-time. 

However, simply stitching these two technologies together creates a conflict, where the AI's "hallucinated" details confuse the traditional codec. Generative models are designed to make images look good to the human eye (Perceptual Quality), often inventing realistic textures like grass or hair that may not pixel-match the original video. Traditional codecs like H.265, however, are mathematically rigid: they expect precise pixel alignment for prediction. If the generative model generates a texture that is slightly offset, traditional codecs interpret this as a massive error and wastes data trying to "fix" it, negating our bitrate savings. To solve this, we leverage the fact that the traditional decoding process is essentially a sequence of differentiable operations. 
This enables us to extend the iterative inversion into an end-to-end optimization loop. We no longer optimize the latent code solely for I-frame reconstruction. Instead, we introduce additional supervision to force the generated I-frame to align with the motion vectors and residuals of the traditional codec. This ensures high perceptual quality across all frames.

We implement \emph{HybridPrompt} on a commercial smartphone and demonstrate that it is the first generative-based system to achieve real-time 1080p decoding. Our contributions are threefold: First, we propose a generic hybrid architecture that leverages generative I-frames to subsidize the bitrate budget of traditional P-frames. Second, we introduce a differentiable bridge that enables end-to-end optimization across the disparate domains of neural generation and traditional prediction, solving the pixel-misalignment problem. Finally, we integrate a two-stage generation strategy. This enables a decoding speed of over 150 FPS on an iPhone 16 Pro Max, surpassing pure neural baselines by orders of magnitude. Simultaneously, we achieve an average LPIPS gain of 8\% over traditional codecs at 200kbps. 

\vspace{-2mm}
\section{Motivation and Key Insights}
\label{sec:motivation}

\subsection{The Dilemma: Quality vs. Efficiency in Mobile VoD}
\label{sec:dilemma}

\begin{figure}[t]
    \centering
    \begin{subfigure}[b]{0.115\textwidth}
        \centering
        \includegraphics[width=\textwidth]{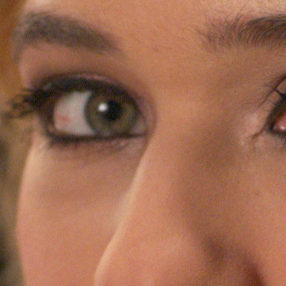}
        \caption{GT}
        \label{fig:gt}
    \end{subfigure}
    \hfill
    \begin{subfigure}[b]{0.115\textwidth}
        \centering
        \includegraphics[width=\textwidth]{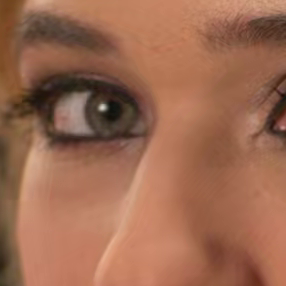}
        \caption{1200 kbps}
        \label{fig:1200kbps}
    \end{subfigure}
    \hfill
    \begin{subfigure}[b]{0.115\textwidth}
        \centering
        \includegraphics[width=\textwidth]{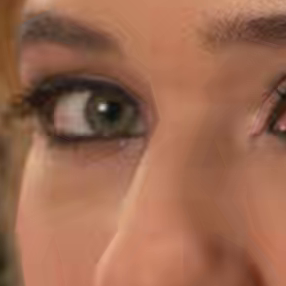}
        \caption{600 kbps}
        \label{fig:600kbps}
    \end{subfigure}
    \hfill
    \begin{subfigure}[b]{0.115\textwidth}
        \centering
        \includegraphics[width=\textwidth]{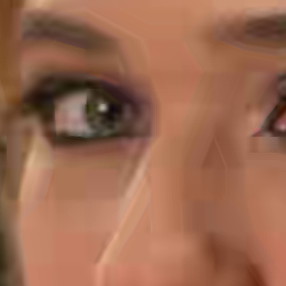}
        \caption{200 kbps}
        \label{fig:200kbps}
    \end{subfigure}
    \vspace{-8mm}
    \caption{We compressed (H.265) a 1080p@30fps video with a group of picture (GOP) length of 120 frames at different bitrates. It can be observed that at lower bitrates, block artifacts and blurring become more pronounced, resulting in noticeably degraded quality.}
    \label{fig:quality_comparison}
    \vspace{-2mm}
\end{figure}

Video on Demand (VoD) on mobile devices faces a fundamental conflict between bandwidth constraints and computational capability. 
Traditional algorithms such as H.264~\cite{264}, H.265~\cite{265}, VP8~\cite{bankoski2011technical}, and VP9~\cite{mukherjee2015technical} are computationally lightweight but struggle to maintain quality under low bandwidth. 
As shown in Figure~\ref{fig:quality_comparison}, when the bitrate is reduced (e.g., to 200 kbps), the quality of I-frames collapses, exhibiting severe block artifacts. Since I-frames serve as references for subsequent P- and B-frames, this degradation propagates through the entire Group of Pictures (GOP), creating a quality bottleneck inherent to the predictive coding paradigm.

\begin{table}[t]
\centering
\caption{Average time required to generate a single 1080p image using Stable Diffusion on several mobile devices. Per-frame generation remains far from real-time on current hardware.}
\vspace{-4mm}
\begin{tabular}{l c}
\toprule
\textbf{Device} & \textbf{Avg. Time per Frame (s)} \\
\midrule
iPhone 12 Pro Max & 9.19 \\
iPhone 13 Pro Max & 8.69 \\
iPhone 14 Pro Max & 4.11 \\
iPhone 16 Pro Max & 2.61 \\
\bottomrule
\end{tabular}
\label{tab:motivation-per_frame_latency}
\vspace{-4mm}
\end{table}

Conversely, emerging neural codecs~\cite{sd, wu2024promptus} offer superior perceptual quality by leveraging generative priors. However, they are computationally prohibitive for mobile devices. As demonstrated in Table~\ref{tab:motivation-per_frame_latency}, generating a single 1080p frame using Stable Diffusion takes over 2 seconds even on an iPhone 16 Pro Max, which is orders of magnitude slower than the 33ms requirement for real-time playback (30 FPS).
This creates a dilemma: we need the visual quality of neural models but the decoding speed of traditional codecs.

\subsection{Feasibility of Hybrid Architecture}
\label{sec:feasibility}

Our first insight is that we can achieve the "best of both worlds" through a hybrid architecture. 
The core observation is that neural codecs are asymmetrically efficient for I-frames. 
As shown in Table~\ref{tab:eval-iframe-quality-size1}, a neural-generated I-frame (Promptus) achieves significantly better perceptual quality (LPIPS 0.463 vs 0.51) with a much smaller file size (8.8 KB vs 11.2 KB) compared to a traditional codec (H.265).

This suggests a strategy of "Bitrate Arbitrage": just using the neural codec for I-frames to save a massive amount of bitrate, and then reallocating this budget to subsequent P-frames encoded by a traditional codec (which is computationally cheap). Theoretically, this should improve the overall quality of the entire video stream.

\begin{table}[t]
\caption{Comparison of I-frame perceptual quality (LPIPS ↓, lower is better) and frame size (KB) between neural and traditional coding under different target bitrates. Neural keyframes consistently achieve better quality-to-size tradeoffs than traditional I-frames.}
\label{tab:eval-iframe-quality-size1}
\centering
\begin{tabular}{lcccc}
\toprule
\textbf{Bitrate (kbps)} & 200 & 400 & 600 & 1200 \\
\midrule
\multicolumn{5}{l}{\textbf{Traditional (H.265)}} \\
LPIPS ↓ & 0.543 & 0.510 & 0.494 & 0.454 \\
Size (KB) & 7.4 & 11.2 & 14.8 & 27.7 \\
\midrule
\multicolumn{5}{l}{\textbf{Neural (Promptus)}} \\
LPIPS ↓ & \multicolumn{2}{c}{0.463} & \multicolumn{2}{c}{0.552} \\
Size (KB) & \multicolumn{2}{c}{8.8} & \multicolumn{2}{c}{4.4} \\
\bottomrule
\end{tabular}
\vspace{-6mm}
\end{table}

\subsection{The Challenge: The Perception-Precision Misalignment}
\label{sec:challenge}

However, a naive combination of Neural and Traditional frames fails in practice. This is due to a fundamental conflict between \textbf{Perceptual Realism} and \textbf{Pixel-wise Precision}.

Generative neural codecs are designed to maximize perceptual quality, reconstructing high-frequency details (e.g., grass textures, hair strands) that appear authentic to the human eye. While these details are visually high-quality, they may not strictly align with the pixel grid of the ground truth. Traditional codecs (like H.265), in contrast, rigorously optimize for objective metrics like MSE. They demand rigid \textbf{pixel-wise precision} and treat even minor spatial misalignments of realistic textures as severe errors.

\begin{figure}[t]
    \centering
    \begin{subfigure}[b]{0.155\textwidth}
        \centering
        \includegraphics[width=\textwidth]{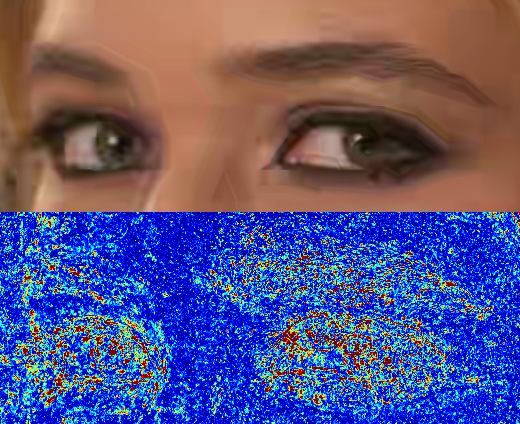}
        \vspace{-4mm}
        \caption{Trad. (H.265)}
        \label{fig:mse_h265}
    \end{subfigure}
    \hfill
    \begin{subfigure}[b]{0.155\textwidth}
        \centering
        \includegraphics[width=\textwidth]{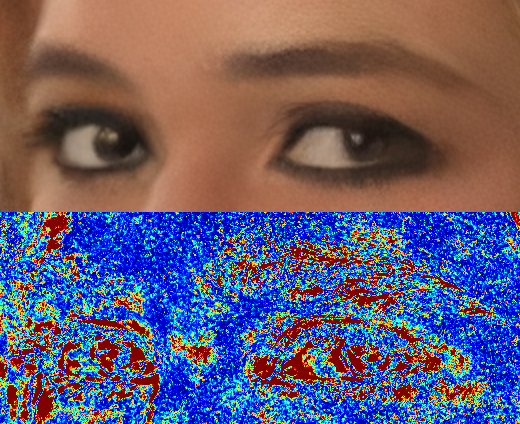}
        \vspace{-4mm}
        \caption{Naive Neural}
        \label{fig:mse_naive}
    \end{subfigure}
    \hfill
    \begin{subfigure}[b]{0.155\textwidth}
        \centering
        \includegraphics[width=\textwidth]{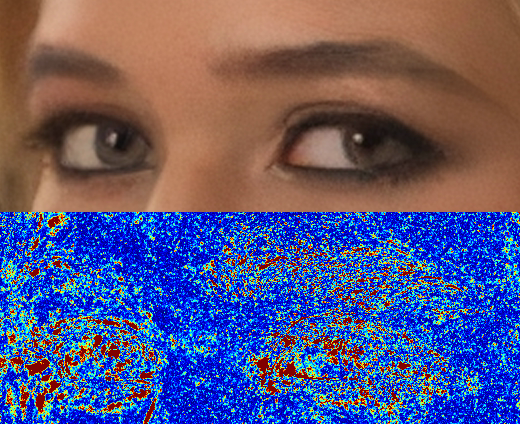}
        \vspace{-4mm}
        \caption{Ours(Optimized)}
        \label{fig:mse_ours}
    \end{subfigure}
    \vspace{-8mm}
    
    \caption{The Perception-Precision Misalignment. All three I-frames are compressed to an identical size of 8.8 KB. Each subfigure displays the \textbf{RGB frame (Top)} and its corresponding \textbf{MSE Heatmap (Bottom)}. }

    \label{fig:mse_comparison}
    \vspace{-4mm}
\end{figure}

We visualize this phenomenon in Figure~\ref{fig:mse_comparison}, where we compare a traditional I-frame and a ``Naive Neural'' I-frame that is directly generated by the model without end to end optimization, both compressed to an identical size of 8.8 KB. 
The Traditional I-frame (a) is blurry but mathematically ``precise'' (low MSE, shown in blue). 
The Naive Neural I-frame (b) is visually sharp and realistic, but mathematically ``misaligned'' (high MSE, shown in red). 
When a neural-generated frame serves as a reference, the traditional codec interprets these visually plausible details as significant error. Consequently, it is forced to expend an excessive amount of bitrate attempting to ``correct'' the pixel-level inconsistencies. This counter-productive correction leads to a \textbf{dual degradation}: it consumes the bitrate savings while simultaneously destroying the sharp neural textures (worsening perceptual quality) and failing to fully restore pixel alignment (worsening MSE).

\begin{figure}[t]
  \centering
  \includegraphics[width=0.9\linewidth]{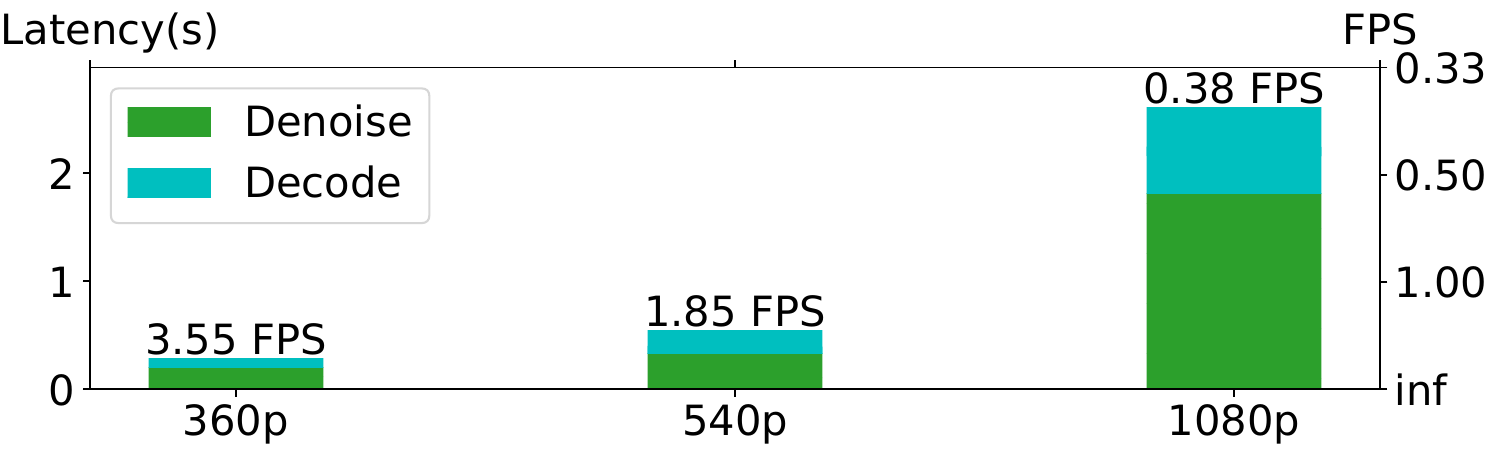}
  \vspace{-2mm}
  \caption{
    Runtime analysis of diffusion-based generation under different resolutions, measured on an iPhone 16 Pro Max. Lower resolutions correspond to significantly reduced latency.
  }
  \label{fig:motivation-latency_resolution_steps}
  \vspace{-4mm}
\end{figure}

\begin{figure*}[t]
  \centering
  \begin{subfigure}[b]{0.245\linewidth}
    \includegraphics[width=\linewidth]{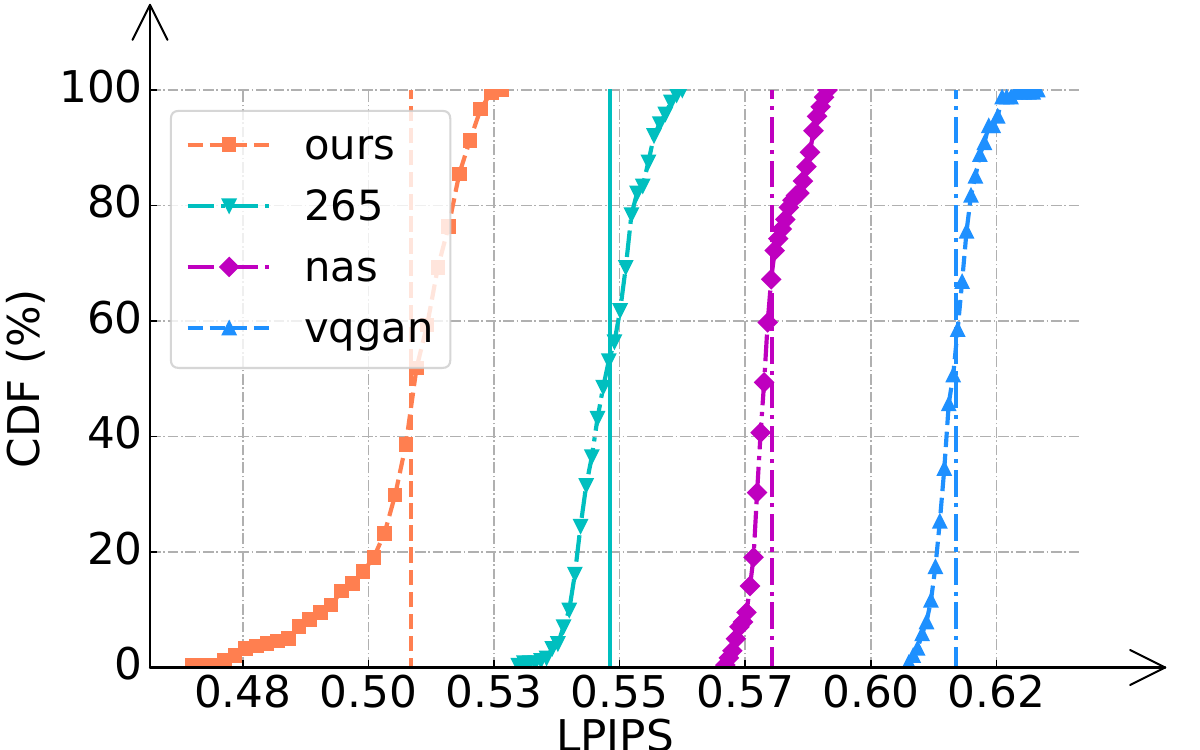}
    \vspace{-6mm}
    \caption{200 kbps}
  \end{subfigure}
  \hfill
  \begin{subfigure}[b]{0.245\linewidth}
    \includegraphics[width=\linewidth]{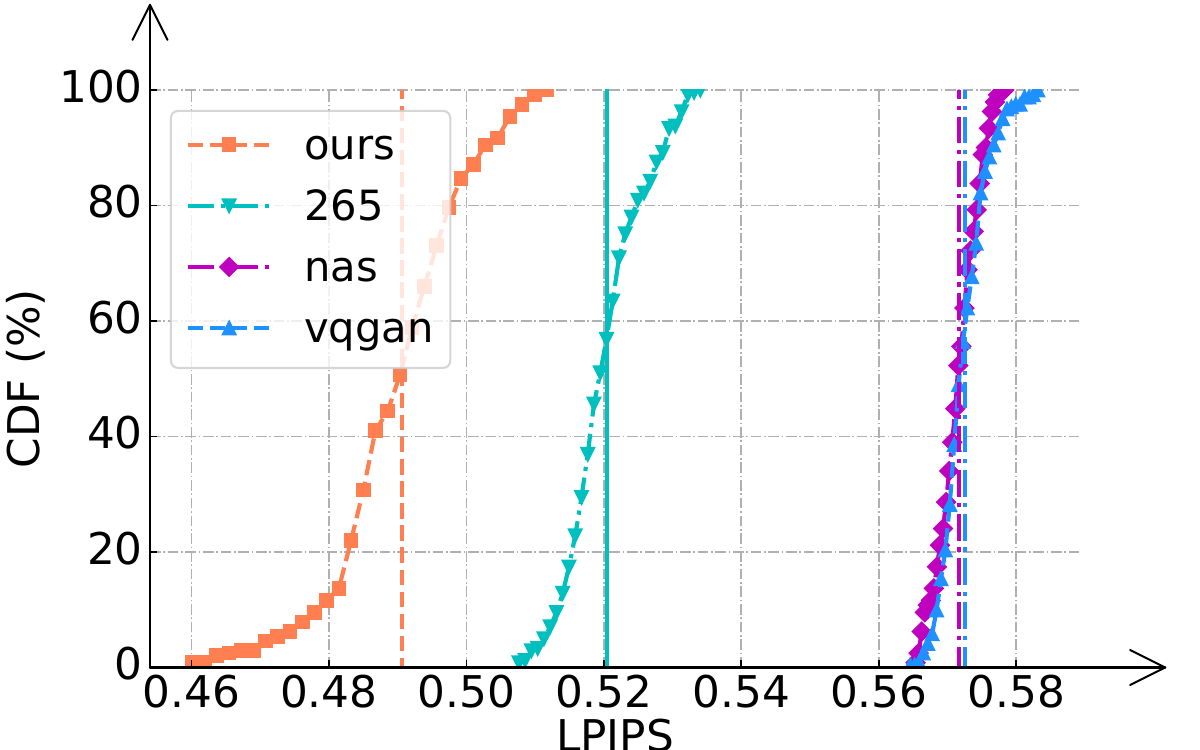}
    \vspace{-6mm}
    \caption{400 kbps}
  \end{subfigure}
  \hfill
  \begin{subfigure}[b]{0.245\linewidth}
    \includegraphics[width=\linewidth]{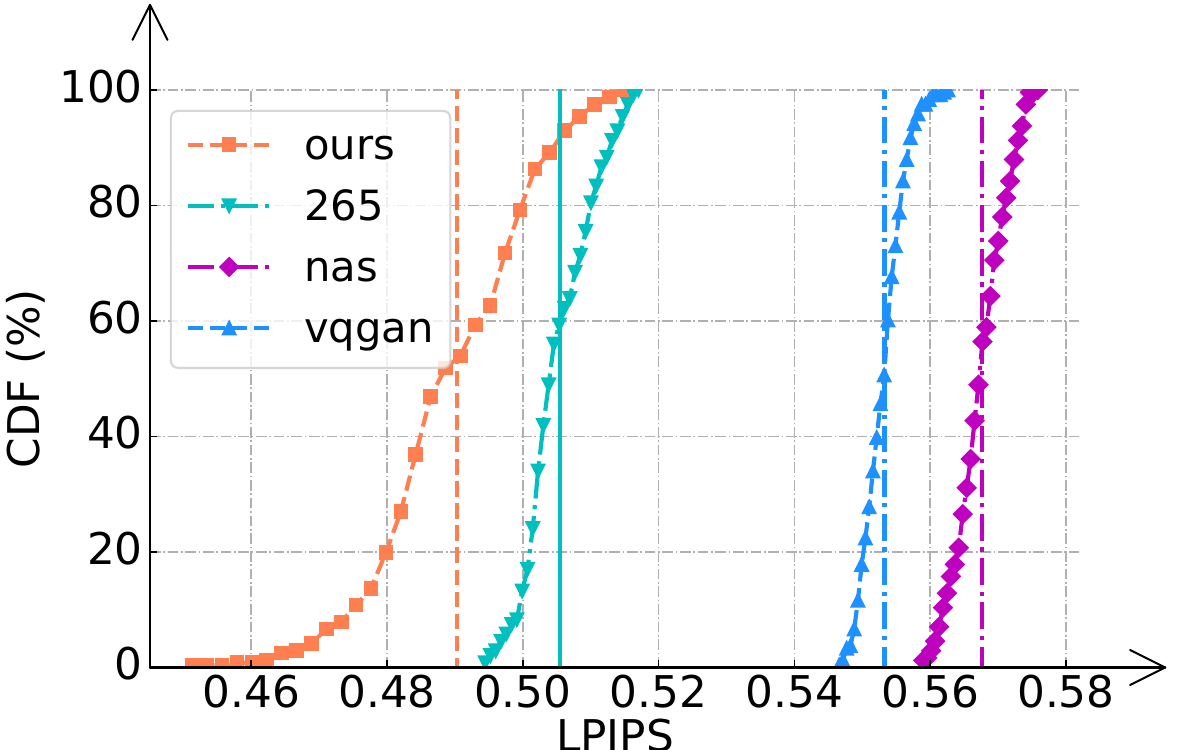}
    \vspace{-6mm}
    \caption{600 kbps}
  \end{subfigure}
  \hfill
  \begin{subfigure}[b]{0.245\linewidth}
    \includegraphics[width=\linewidth]{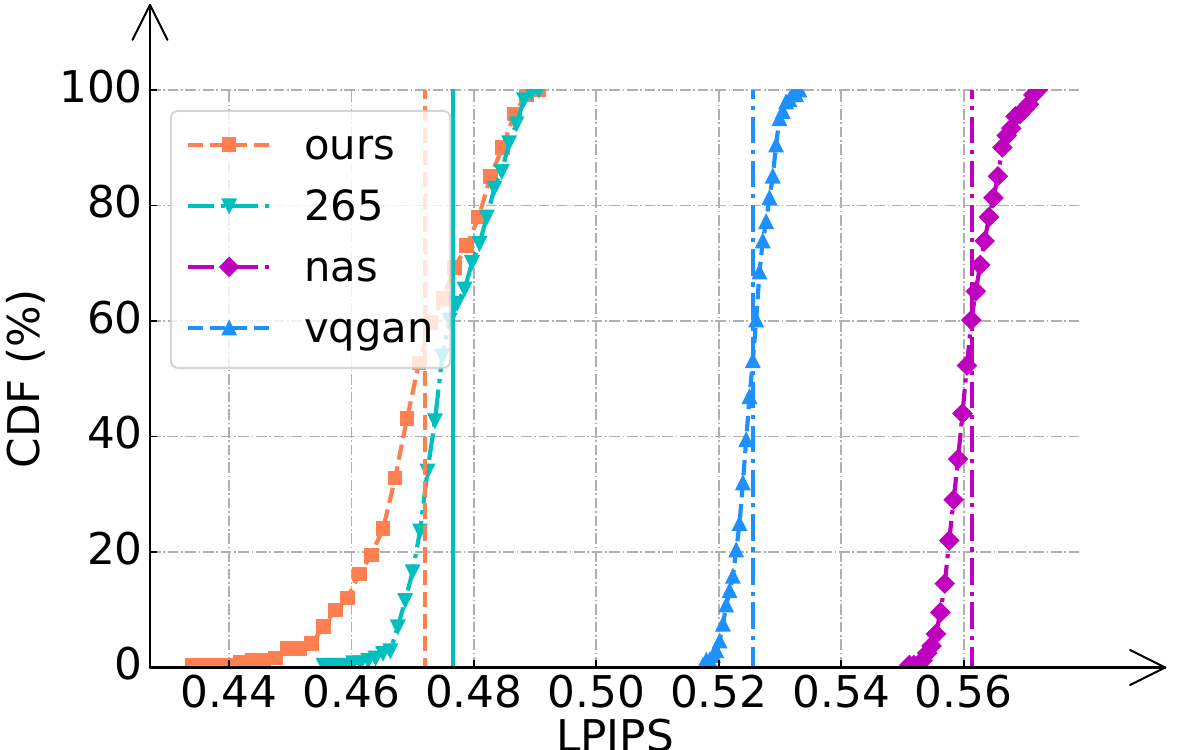}
    \vspace{-6mm}
    \caption{1200 kbps}
  \end{subfigure}
  \vspace{-4mm}
  \caption{Perceptual quality (LPIPS, lower is better) comparison across different bitrate settings at 1080p resolution. HybridPrompt consistently achieves the lowest LPIPS (best quality) under all bandwidth conditions.}
  \Description{CDF curves comparing HybridPrompt with baselines on the UVG dataset.
  The x-axis is LPIPS, where lower values indicate better perceptual similarity.
  HybridPrompt shows lower LPIPS than baselines across 200--1200 kbps, with about 0.04 absolute improvement at low bitrates.}
  \label{fig:eval-bitrate_lpips_all}
  \vspace{-3mm}
\end{figure*}

\subsection{Insight: Bridging the Gap via Differentiable End-to-End Optimization}
\label{sec:insight_optimization}

To resolve this conflict, we need the generative model to accommodate the traditional codec. 
Our key insight is that the entire traditional decoding sequence is composed of differentiable mathematical operations.

Specifically, once the motion vectors and residuals are obtained, the reconstruction process consists of three distinct steps.
First, the reference frame undergoes \textbf{warping} based on motion vectors, which functions as a differentiable sampling operation.
Second, the decoded \textbf{residuals are added} to the warped frame, which is a differentiable addition operation. 
Third, the frame passes through \textbf{deblocking filters}, which can be modeled as differentiable convolutions.

Since every step in this chain is differentiable, we can construct an \textbf{end-to-end optimization loop}.
Instead of optimizing the I-frame solely for its own visual quality, we optimize its latent representation to minimize the reconstruction error of all subsequent P- and B-frames.
By backpropagating the loss from these frames through the warping, adding and filtering layers to the I-frame latent, we force the generative model to produce details that are not only visually realistic but also \textbf{friendly for traditional codecs}. 

As shown in Figure~\ref{fig:mse_comparison}(c), our optimized I-frame retains the sharp textures of the neural model. 
Crucially, it significantly suppresses the pixel-level error compared to the naive neural approach. 
While the MSE does not strictly surpass H.265, this improved alignment prevents the traditional codec from wasting bitrate on unnecessary corrections, thereby restoring the efficiency of the hybrid stream.


\subsection{Speed Optimization: Decoupling Resolution for Real-Time Decoding}
\label{sec:speed}

With the quality and compatibility issues resolved, the final hurdle is speed. 
We conducted experiments on an iPhone 16 Pro Max to measure the generation latency of diffusion models across various resolutions. 
As shown in Figure~\ref{fig:motivation-latency_resolution_steps}, our results indicate that the inference time is approximately proportional to the total pixel count, meaning lower resolutions yield significantly faster generation.

Leveraging this observation, we adopt a \textbf{Two-stage Generation} strategy to achieve real-time performance. 
Instead of directly generating full 1080p frames, we first generate a 540p low-resolution image. 
This image is subsequently upsampled to 1080p via a lightweight super-resolution module. 
Crucially, since this upsampling operation is fully differentiable, we integrate it into our end-to-end training pipeline. 
This joint optimization minimizes the quality degradation typically associated with super-resolution, ensuring the final high-resolution output maintains high fidelity.

\section{System Design}
\label{sec:method}


\subsection{Architecture Overview}
As Fig. ~\ref{fig:overview}, the lifecycle proceeds as follows:

\noindent\textbf{Server-Side Transcoding.} The server takes raw video as input and decomposes it into Groups of Pictures (GOPs). For each GOP, the first frame is inverted into a compact \textit{generative latent code} via our neural pipeline. The subsequent frames are encoded using a standard traditional H.265 encoder, but with the neural-generated I-frame as the reference.

\vspace{0.5em}
\noindent\textbf{Bitstream Muxing.} We devise a custom transport format. Instead of transmitting pure bulky H.265 bitstream, we discard the I-frame and inject the compact latent code into the container. The final hybrid bitstream consists of two interleaved tracks: the \textit{Neural Stream} of I-frames and the \textit{Legacy Stream} (Motion Vectors and Residuals) of P-/B-frames.

\vspace{0.5em}
\noindent\textbf{Client-Side Dual-Decoding.} Upon receiving the stream, the client demuxes the data. The latent codes are decoded by the Neural Engine to reconstruct the I-frames, which are then placed into the playback buffer to serve as reference frames for the hardware decoder to reconstruct the P/B-frames.

\vspace{-3mm}
\subsection{Server-Side: Transcoding Pipeline}
\begin{figure}[t]
  \centering
  \includegraphics[width=0.85\linewidth]{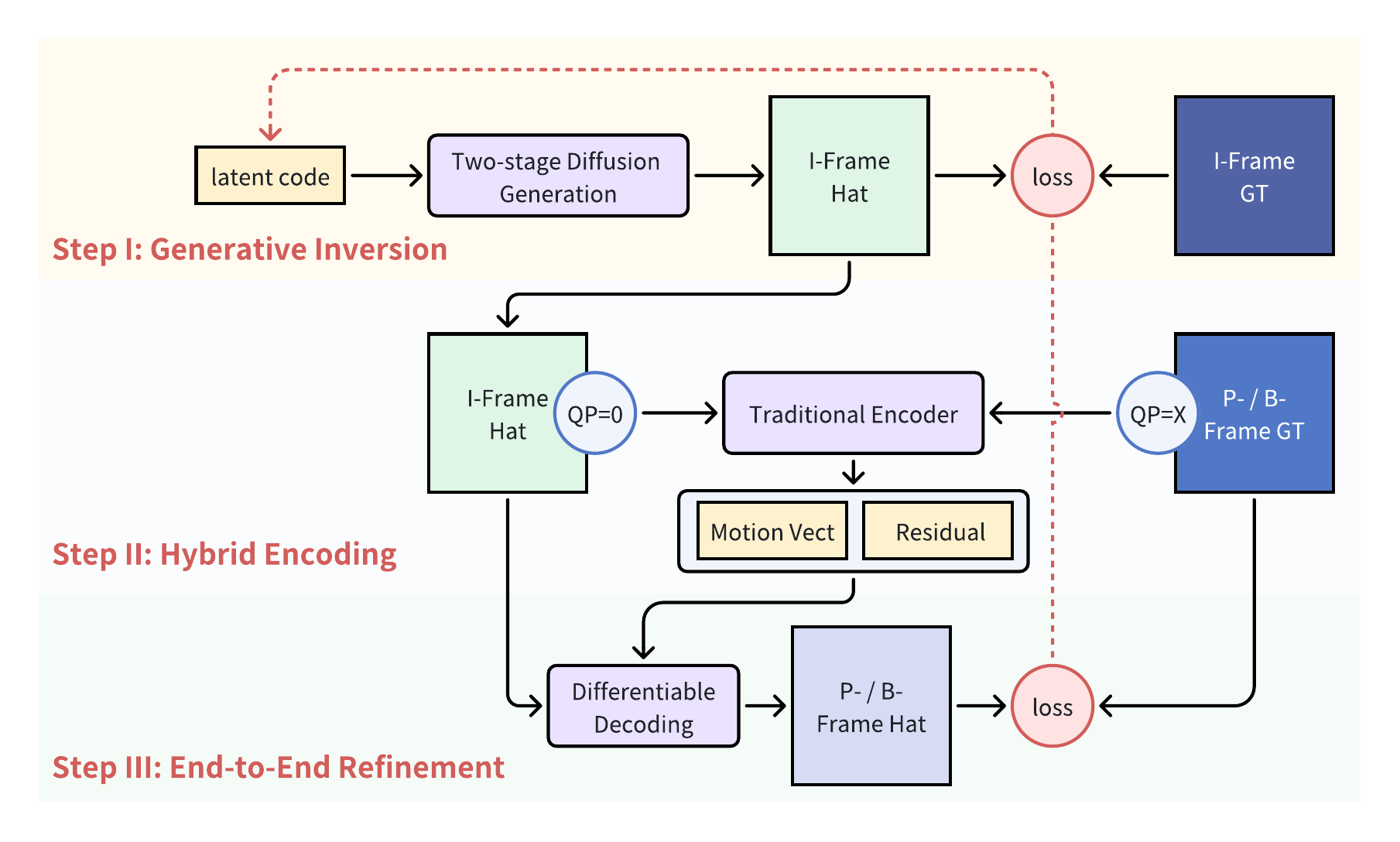}
  \vspace{-4mm}
  \caption{Overview of the Hybrid Transcoding Pipeline.}
  \label{fig:pipeline}
  \vspace{-4mm}
\end{figure}

As shown in Figure~\ref{fig:pipeline}, the core contribution of our method is the server-side optimization strategy, which ensures that the generative I-frames are not only visually pleasing but also mathematically consistent with the H.265 prediction logic. This process involves three sequential steps below:

\vspace{0.5em}
\noindent\textbf{Step I: Generative Inversion.} 
First, for each raw I-frame $I_{gt}$, we aim to find a latent code $z$ that can reconstruct it. To ensure real-time performance on mobile clients, we employ a \textit{two-stage generation strategy}. The code $z$ first drives a diffusion model to synthesize a low-resolution(540p) image, which is then upsampled to 1080p via a differentiable super-resolution module. We iteratively optimize $z$ to minimize the perceptual difference between the generated frame $\hat{I}_{gen}$ and $I_{gt}$, initializing the visual "anchor" for the GOP.

\vspace{0.5em}
\noindent\textbf{Step II: Hybrid Motion Encoding \& Rate Control.} 
Once $\hat{I}_{gen}$ is generated, we then encode the remaining frames in the GOP using a traditional H.265 encoder. A critical challenge is ensuring the codec uses the \textit{exact} artifacts of $\hat{I}_{gen}$ for motion estimation. To address this, we implement a \textit{Reference Injection with QP=0} strategy, where we feed $\hat{I}_{gen}$ into the H.265 encoder and force the Quantization Parameter (QP) to \textbf{0} (lossless mode) for the I-frame. This ensures the encoder calculates Motion Vectors (MVs) and Residuals based exactly on our generative output without introducing additional compression noise. Note that the resulting massive H.265 I-frame bitstream is discarded, and only the P/B-frame data is kept. To manage bandwidth, we simultaneously apply \textit{Binary Search Allocation}, dynamically adjusting the QP for the P/B-frames until the total bitrate (Size of Code $z$ + Size of P/B Stream) meets the target bandwidth budget.

\vspace{0.5em}
\noindent\textbf{Step III: End-to-End Joint Refinement.} 
The initial code $z$ from Step I is optimized solely for the I-frame, which may yield suboptimal prediction for the rest of the GOP. In this final stage, we freeze the MVs and Residuals obtained from Step II and perform end-to-end fine-tuning. We construct a differentiable decoding chain where predicted frames are reconstructed via warping $\hat{I}_{gen}$ using the fixed MVs. We calculate the reconstruction loss across \textit{all} frames in the GOP and backpropagate gradients to update the latent code $z$. This process forces the generative I-frame to align with the motion trajectory of the video, ensuring that the "hallucinated" details are compatible with the codec's prediction logic. 

We do notice that the decoding chain contains some non-smooth or gradient-attenuating operators (such as clipping, rounding, or threshold-like steps). But they do not fully block gradients everywhere. As shown in ~\ref{fig:mse_comparison}(c), gradients flowing through the majority of differentiable components are sufficient for end-to-end refinement to work and to make $\hat{I}_{gen}$ compatible with the H.265 prediction logic.


\subsection{Client-Side: Parallel Decoding}
To achieve real-time playback on a real phone, we implement a parallelized decoding pipeline. The process is asynchronous: while the Video Processing Unit (VPU) is decoding the P- and B-frames of GOP $N$, the Neural Processing Unit (NPU) is generating the I-frame for GOP $N+1$ in the background.

\vspace{0.5em}
\noindent\textbf{Just-in-Time Stream Stitching.} 
To ensure seamless integration with commercial hardware decoders without requiring driver-level modifications, we introduce a lightweight re-encoding step. Once the I-frame is generated by the NPU, we immediately perform a software encoding at \textbf{QP=0} (lossless) and concatenate this local I-frame header with the received P/B-frame bitstream. This on-the-fly stitching constructs a standard-compliant H.265 bitstream, allowing the VPU to treat the hybrid stream as a normal video file. Our experiments demonstrate that this intermediate re-encoding is extremely lightweight, introducing negligible overhead and strictly maintaining the real-time decoding speed.
\begin{figure}[t]
  \centering
  \includegraphics[width=0.35\textwidth]{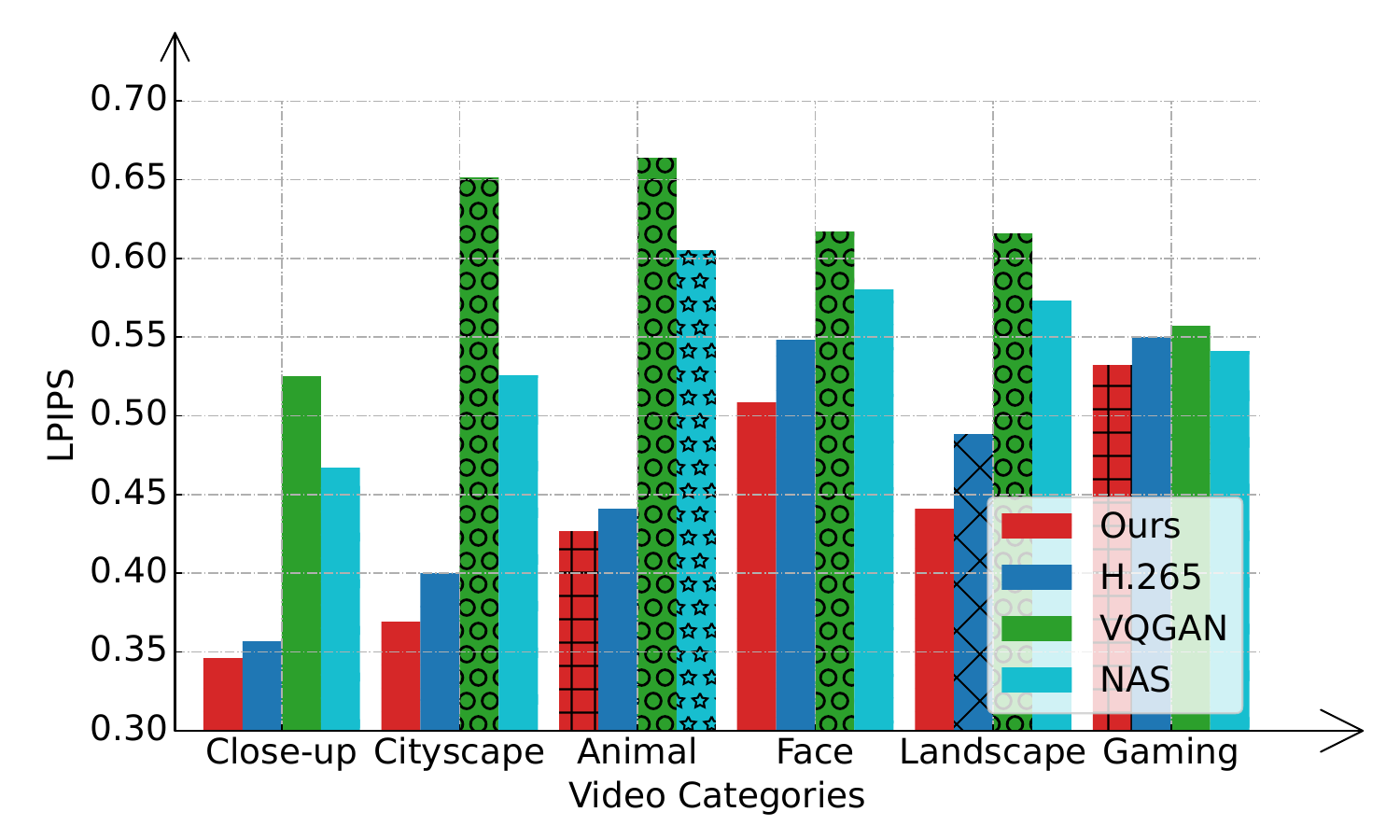}
  \vspace{-4mm}
  \caption{LPIPS (lower is better) comparison across diverse video categories from the UVG dataset.}
  \label{fig:eval-lpips_generalization}
  \vspace{-4mm}
\end{figure}

\section{Evaluation}
\label{sec:experiment}


We design our evaluation to answer three pivotal questions: (1) \textbf{Quality \& Generalization:} Does the hybrid architecture consistently outperform baselines across varying bandwidths and diverse content scenarios? (2) \textbf{Mechanism:} How does the ``Bitrate Arbitrage'' strategy effectively reallocate bandwidth to enhance overall fidelity? (3) \textbf{Efficiency:} Can the system achieve real-time performance on mobile hardware, and what is the contribution of each architectural component to the speed-quality trade-off?

\subsection{Experimental Setup}

\noindent\textbf{Testbed.} 
The server-side transcoding is performed on a workstation equipped with an NVIDIA RTX 4090 GPU (24 GB VRAM). 
For the client side, we deploy our decoding pipeline on an iPhone 16 Pro Max (A18 Pro chip). We leverage the Neural Engine (NPU) for the generative inference and the hardware Video Processing Unit (VPU) for H.265 decoding. 

\noindent\textbf{Datasets \& Metrics.} 
We use the \textbf{UVG dataset}~\cite{mercat2020uvg}, standardized to 1080p at 30 FPS. This dataset covers diverse categories including \textit{Cityscapes, Human Faces}, and \textit{3D Gaming}.
To evaluate perceptual quality, we report LPIPS~\cite{zhang2018unreasonable} (lower is better), as it aligns more closely with human vision than PSNR for generative content. We also measure Frames Per Second (FPS) to quantify throughput.

\noindent\textbf{Baselines.} We compare our approach against three distinct categories of methods. First, representing \textbf{Traditional Codecs}, we evaluate H.265~\cite{265}, the current industrial standard. Second, for \textbf{Neural-enhanced Streaming} (Super-Resolution), we evaluate NAS~\cite{yeo2018neural}, approximating the mobile and real-time version of NEMO~\cite{yeo2020nemo}, to benchmark against content-aware super-resolution methods~\cite{park2023omnilive,zhou2022cadm}. Third, representing \textbf{Lightweight Generative Neural Codecs}, we select VQGAN~\cite{esser2021taming,li2023reparo}, a token-based approach that quantizes latent variables~\cite{van2017neural} via a VAE~\cite{kingma2014auto}.


\begin{table}[h]
\caption{Comparison of P/B-frame sizes. Our method allocates significantly more bandwidth to P/B frames compared to standard H.265, resulting in better quality.}
\vspace{-2mm}
\label{tab:eval-bpframe-quality-size}
\centering
\small
\begin{tabular}{lcccc}
\toprule
\textbf{\makecell{Target\\Bitrate}} & \multicolumn{2}{c}{\textbf{Ours (Hybrid)}} & \multicolumn{2}{c}{\textbf{Standard H.265}} \\
\cmidrule(r){2-3} \cmidrule(l){4-5}
 & LPIPS $\downarrow$ & Frame Size & LPIPS $\downarrow$ & Frame Size \\
\midrule
200 kbps (P)   & \textbf{0.475} & \textbf{2.2 KB} & 0.549 & 1.6 KB \\
200 kbps (B)   & \textbf{0.474} & \textbf{0.4 KB} & 0.536 & 0.25 KB \\
\midrule
600 kbps (P)   & \textbf{0.451} & \textbf{8.6 KB} & 0.494 & 4.4 KB \\
600 kbps (B)   & \textbf{0.450} & \textbf{1.7 KB} & 0.494 & 0.9 KB \\
\midrule
1200 kbps (P)  & \textbf{0.462} & \textbf{19.9 KB} & 0.467 & 9.6 KB \\
1200 kbps (B)  & \textbf{0.440} & \textbf{4.1 KB} & 0.460 & 1.7 KB \\
\bottomrule
\end{tabular}
\vspace{-4mm}
\end{table}

\subsection{Perceptual Quality Analysis}

\noindent\textbf{Overall Quality \& Generalization.} 
We evaluate whether our hybrid architecture delivers superior perceptual quality across varying network conditions and content types. 
Figure~\ref{fig:eval-bitrate_lpips_all} presents the Cumulative Distribution Function (CDF) of LPIPS scores for all frames at four bitrate levels. Our method consistently achieves the best quality distribution compared to H.265, NAS, and VQGAN. This advantage is particularly pronounced at 200 kbps, where traditional codecs suffer from severe degradation.
Furthermore, we validate robustness across diverse scenarios, including \textit{Close-ups}, \textit{Cityscapes}, and \textit{High-dynamic Gaming}, in Figure~\ref{fig:eval-lpips_generalization}. Our approach demonstrates universal applicability, consistently outperforming baselines regardless of scene complexity. 

\vspace{0.5em}
\noindent\textbf{Mechanism: The "Bitrate Arbitrage" Effect.} 
The superior quality of our P/B-frames, despite using the same H.265 backend, is explained by resource reallocation. 
As analyzed in Table~\ref{tab:eval-bpframe-quality-size}, our generative I-frame representation is extremely compact, allowing us to "steal" bandwidth from I-frames and reallocate it to P- and B-frames. 
For instance, at a target of 600 kbps, our method allocates an average of \textbf{8.6 KB} to P-frames, nearly double the \textbf{4.4 KB} allocated by standard H.265. This "Bitrate Arbitrage" ensures that motion vectors and residuals are encoded with significantly higher fidelity, resulting in a sharper overall video stream.

\subsection{System Efficiency \& Ablation Study}

To quantify the contribution of individual system components, we conduct an ablation study on the iPhone 16 Pro Max, with results summarized in Table~\ref{tab:ablation}.

\noindent\textbf{Real-Time Performance.} 
The full HybridPrompt system achieves a remarkable decoding speed of \textbf{153.7 FPS}. This confirms that our parallelized architecture that offloads I-frames to the NPU and handles P-frames on the VPU effectively eliminates the generative bottleneck. 

\noindent\textbf{Component Analysis.} 
We further analyze the impact of specific modules. First, the \textbf{Two-Stage Generation} strategy is critical for efficiency; replacing it with direct high-resolution generation drops throughput to 41.8 FPS, significantly increasing power consumption. Second, the \textbf{Joint Optimization} is essential for fidelity; disabling the end-to-end refinement degrades LPIPS from 0.458 to 0.463. This validates that simply ``gluing'' a generator to a codec is insufficient and gradient flow through the codec is necessary to strictly align the generative prior with the video stream. Finally, although purely neural encoding of every frame yields superior quality, the resulting latency is prohibitive, rendering it completely unacceptable for mobile deployment. 

\begin{table}[h]
\centering
\caption{Ablation Study on iPhone 16 Pro Max.}
\vspace{-3mm}
\label{tab:ablation}
\begin{tabular}{lcc}
\toprule
\textbf{Configuration} & \textbf{LPIPS ($\downarrow$)} & \textbf{FPS ($\uparrow$)} \\
\midrule
\textbf{Full HybridPrompt (Ours)}         & \textbf{0.458} & \textbf{153.7} \\
w/o Joint Frame+Bitrate Opt.     & 0.463 & 153.7 \\
w/o Two-Stage Generation         & 0.454 & 41.8 \\
w/o Hybrid Strategy (Prompt-Only)   & 0.384 & 0.38 \\
\bottomrule
\end{tabular}
\vspace{-6mm}
\end{table}
\section{Conclusion and Future Work}
In this work, we presented HybridPrompt, a mobile-friendly video coding framework that leverages generative priors to achieve superior perceptual quality at low bitrates. But several avenues remain for exploration:

\noindent \textbf{Adaptive Bitrate Allocation.} Our system now uses a heuristic strategy to balance the bandwidth between the neural stream and the traditional stream. Future work will explore better rate control to dynamically optimize the trade-off between semantic I-frames and texture-rich P-frames based on real-time channel fluctuations and scene complexity. 

\noindent \textbf{Next-Generation Hybrid Coding.} While we demonstrated effectiveness with H.265, the hybrid paradigm is codec agnostic. Future iterations will integrate more advanced traditional codecs, such as AV1 or VVC, and more powerful diffusion-based models to further improve compression efficiency. Additionally, we aim to explore quantization techniques to further reduce the energy footprint on mobile NPUs.

This work was sponsored by the NSFC grant(62431017), Bytedance Grant(CT20241126107484). We gratefully acknowledge the support of State Key Laboratory of Media Convergence Production Technology and Systems, Key Laboratory of Intelligent Press Media Technology. The corresponding author is Xinggong Zhang(zhangxg@pku.edu.cn). 
\bibliographystyle{ACM-Reference-Format}
\bibliography{aaai2026}

@String{Computing = "Computing" }

@String{Computer = "{IEEE} Computer" }

@misc{264,
    title={H.264},
    note={\url{https://www.itu.int/rec/T-REC-H.264}},
    year={2024},
}

@misc{265,
    title={H.265},
    note={\url{https://www.itu.int/rec/T-REC-H.265}},
    year={2024},
}

@inproceedings{bankoski2011technical,
  title={Technical overview of VP8, an open source video codec for the web},
  author={Bankoski, Jim and Wilkins, Paul and Xu, Yaowu},
  booktitle={2011 IEEE International Conference on Multimedia and Expo},
  pages={1--6},
  year={2011},
  organization={IEEE}
}

@article{mukherjee2015technical,
  title={A technical overview of vp9—the latest open-source video codec},
  author={Mukherjee, Debargha and Han, Jingning and Bankoski, Jim and Bultje, Ronald and Grange, Adrian and Koleszar, John and Wilkins, Paul and Xu, Yaowu},
  journal={SMPTE Motion Imaging Journal},
  volume={124},
  number={1},
  pages={44--54},
  year={2015},
  publisher={SMPTE}
}

@article{li2023reparo,
  title={Reparo: Loss-resilient generative codec for video conferencing},
  author={Li, Tianhong and Sivaraman, Vibhaalakshmi and Fan, Lijie and Alizadeh, Mohammad and Katabi, Dina},
  journal={arXiv preprint arXiv:2305.14135},
  year={2023}
}

@article{kingma2014auto,
  title={Auto-Encoding Variational Bayes},
  author={Kingma, Diederik P and Welling, Max},
  journal={Journal of Machine Learning Research (JMLR)},
  volume={15},
  number={1},
  pages={1929--1958},
  year={2014}
}

@inproceedings{yeo2018neural,
  title={Neural adaptive content-aware internet video delivery},
  author={Yeo, Hyunho and Jung, Youngmok and Kim, Jaehong and Shin, Jinwoo and Han, Dongsu},
  booktitle={13th USENIX Symposium on Operating Systems Design and Implementation (OSDI 18)},
  pages={645--661},
  year={2018}
}

@inproceedings{park2023omnilive,
  title={Omnilive: Super-resolution enhanced 360 video live streaming for mobile devices},
  author={Park, Seonghoon and Cho, Yeonwoo and Jun, Hyungchol and Lee, Jeho and Cha, Hojung},
  booktitle={Proceedings of the 21st Annual International Conference on Mobile Systems, Applications and Services},
  pages={261--274},
  year={2023}
}

@article{zhou2022cadm,
  title={Cadm: Codec-aware diffusion modeling for neural-enhanced video streaming},
  author={Zhou, Qihua and Li, Ruibin and Guo, Song and Dong, Peiran and Liu, Yi and Guo, Jingcai and Xu, Zhenda},
  journal={arXiv preprint arXiv:2211.08428},
  year={2022}
}

@inproceedings{wu2024promptus,
  title={Promptus: Can Prompts Streaming Replace Video Streaming},
  author={Wu, Jiangkai and Liu, Liming and Tan, Yunpeng and Hao, Junlin and Zhang, Liang and Zhang, Xinggong},
  booktitle={Proceedings of the AAAI Conference on Artificial Intelligence},
  year={2026}
}

@inproceedings{esser2021taming,
  title={Taming transformers for high-resolution image synthesis},
  author={Esser, Patrick and Rombach, Robin and Ommer, Bjorn},
  booktitle={Proceedings of the IEEE/CVF conference on computer vision and pattern recognition},
  pages={12873--12883},
  year={2021}
}

@inproceedings{zhang2018unreasonable,
  title={The unreasonable effectiveness of deep features as a perceptual metric},
  author={Zhang, Richard and Isola, Phillip and Efros, Alexei A and Shechtman, Eli and Wang, Oliver},
  booktitle={Proceedings of the IEEE conference on computer vision and pattern recognition},
  pages={586--595},
  year={2018}
}

@misc{global-phenomena24,
    title={2024 Global Internet Phenomena Report},
    author={Sandvine}, 
    note={\url{https://www.sandvine.com/global-internet-phenomena-report-2024}},
    year={2024},
}

@misc{sd,
    title={Stable Diffusion},
    note={\url{https://stability.ai/}},
    year={2024},
}

@inproceedings{ebenezer2022subjective,
  title={Subjective and objective quality assessment of high-motion sports videos at low-bitrates},
  author={Ebenezer, Joshua P and Chen, Yixu and Wu, Yongjun and Wei, Hai and Sethuraman, Sriram},
  booktitle={2022 IEEE International Conference on Image Processing (ICIP)},
  pages={521--525},
  year={2022},
  organization={IEEE}
}

@inproceedings{mercat2020uvg,
  title={UVG dataset: 50/120fps 4K sequences for video codec analysis and development},
  author={Mercat, Alexandre and Viitanen, Marko and Vanne, Jarno},
  booktitle={Proceedings of the 11th ACM multimedia systems conference},
  pages={297--302},
  year={2020}
}

@article{van2017neural,
  title={Neural discrete representation learning},
  author={Van Den Oord, Aaron and Vinyals, Oriol and others},
  journal={Advances in neural information processing systems},
  volume={30},
  year={2017}
}

@inproceedings{yeo2020nemo,
  title={Nemo: enabling neural-enhanced video streaming on commodity mobile devices},
  author={Yeo, Hyunho and Chong, Chan Ju and Jung, Youngmok and Ye, Juncheol and Han, Dongsu},
  booktitle={Proceedings of the 26th Annual International Conference on Mobile Computing and Networking},
  pages={1--14},
  year={2020}
}

\end{document}